# Discovery of TeV Gamma-Ray Emission from the Cygnus Region of the Galaxy

A. A. Abdo,[1] B. Allen,[2] D. Berley,[3] E. Blaufuss,[3] S. Casanova,[4] C. Chen,[2] D. G. Coyne,[5] R.S. Delay,[2] B. L. Dingus,[4] R. W. Ellsworth,[6] L. Fleysher,[7] R. Fleysher,[7] M. M. Gonzalez,[8] J. A. Goodman,[3] E. Hays,[3] C. M. Hoffman,[4] Brian E. Kolterman,[7] L. A. Kelley,[5] C. P. Lansdell,[3] J. T. Linnemann,[1] J. E. McEnery,[9] A. I. Mincer,[7] I. V. Moskalenko,[10] P. Nemethy,[7] D. Noyes,[3] J. M. Ryan,[11] F. W. Samuelson,[4] P. M. Saz Parkinson,[5] M. Schneider,[5] A. Shoup,[12] G. Sinnis,[4] A. J. Smith,[3] A. W. Strong,[13] G. W. Sullivan,[3] V. Vasileiou,[3] G. P. Walker,[4] D. A. Williams,[5] X. W. Xu[4] and G. B. Yodh[2]

[1]Michigan State University, East Lansing, MI
[2]University of California, Irvine, CA
[3]University of Maryland, College Park, MD
[4]Los Alamos National Laboratory, Los Alamos, NM
[5]University of California, Santa Cruz, CA
[6]George Mason University, Fairfax, VA
[7]New York University, New York, NY
[8]Instituto de Astronomia, Universidad Nacional Autonoma de Mexico, D.F., Mexico
[9]NASA Goddard Space Flight Center, Greenbelt, MD
[10]Stanford University, Stanford, CA
[11]University of New Hampshire, Durham, NH
[12]Ohio State University, Lima, OH
[13]Max Planck Institut fur Extraterrestrische Physik, Garching, Germany

Abstract

The diffuse gamma radiation arising from the interaction of cosmic ray particles with matter and radiation in the Galaxy is one of the few probes available to study the origin of the cosmic rays. Milagro is a water Cherenkov detector that continuously views the entire overhead sky. The large field-of-view combined with the long observation time makes Milagro the most sensitive instrument available for the study of large, low surface brightness sources such as the diffuse gamma radiation arising from interactions of cosmic radiation with interstellar matter. In this paper we present spatial and flux measurements of TeV gamma-ray emission from the Cygnus Region. The TeV image shows at least one new source MGRO J2019+37 as well as correlations with the matter density in the region as would be expected from cosmic-ray proton interactions. However, the TeV gamma-ray flux as measured at ~12 TeV from the Cygnus region (after excluding MGRO J2019+37) exceeds that predicted from a conventional model of cosmic ray production and propagation. This observation indicates the existence of either hard-spectrum cosmic-ray sources and/or other sources of TeV gamma rays in the region.

## 1. Introduction

A diffuse gamma-ray flux is expected from the plane of the Galaxy due to cosmic ray interactions with matter and radiation fields. At energies above 1 GeV, the flux from the Galactic plane, as determined from EGRET observations, exceeded predictions based upon the matter density measured in the region and the cosmic-ray spectrum and intensity measured at Earth (Hunter *et al.* 1997, Strong *et al*. 2004). Explanations of this excess include unresolved sources, a varying cosmic ray spectrum or intensity across the Galaxy (Strong *et al.* 2004, Gralewicz *et al.* 1997), a hard electron spectrum leading to an increase in the inverse Compton component (Porter *et al*. 1997), or the addition of a new production mechanism such as the annihilation of relic dark matter particles (de Boer *et al*. 2005).

At higher energies, TeV emission from the Galactic plane was discovered with the data from the Milagro gamma-ray observatory (Atkins *et al.* 2005). While the Milagro flux estimate was consistent with a simple power-law extrapolation from the EGRET data, it has been pointed out that this implies that there exists a TeV excess of diffuse gamma radiation (Prodanovic *et al.* 2006). The HESS telescope has measured the energy spectrum of gamma rays arising from the interactions of cosmic rays near the Galactic Center with a complex of molecular clouds (Aharonian *et al.* 2006). The differential energy spectrum measured by HESS is $E^{-2.3}$ and the flux is a factor of 3-9 times higher than expected based on the local cosmic-ray spectrum and intensity, supporting the hypothesis that the cosmic-ray spectrum depends upon the location in the Galaxy.

The Milagro diffuse flux was measured from the Galactic longitude of 40-100 degrees. The strongest excess, as compared to the EGRET measurements in the same region, was near longitude 80 degrees – the Cygnus Region of the Galaxy. This region of the Galaxy is a natural laboratory for the study of cosmic ray origins. It contains a large column density of interstellar gas that should lead to strong emission of diffuse gamma rays and is also the home of potential cosmic-ray acceleration sites--Wolf-Rayet stars (van der Hucht 2001), OB associations (Bocharev *et al.* 1985), and supernova remnants (Green 2004). The Tibet Air Shower detector also recently reported an excess in the cosmic ray flux from this region (Amenomori *et al.* 2006). In this paper, we report on the results of a detailed analysis of the Milagro data (with substantially higher sensitivity than our previous work) from the Cygnus Region.

**2. Data Analysis**

The Milagro gamma-ray observatory is a water Cherenkov detector that continuously views the entire overhead sky (Sullivan 2001). The large field-of-view combined with the long observation time makes Milagro the most sensitive instrument available for the study of large, low surface brightness sources such as the diffuse gamma radiation arising from interactions of cosmic radiation with interstellar matter. Milagro consists of a 24 million liter water reservoir instrumented with 723 photomultiplier tubes (PMTs) surrounded by an array of 175 water tanks (outriggers). The top layer of 450 PMTs is under 1.3 meters of water and is used to trigger the detector and to reconstruct the direction of the primary gamma ray (or cosmic ray). The bottom layer of 273 PMTs is under 6 meters of water and is used to measure the penetrating component of air showers induced by hadronic cosmic rays. The PMT spacing in the reservoir is 2.7 meters and the area enclosed is 4800 $m^2$. The surrounding array of water tanks is dispersed over 34,000 $m^2$. Each tank is cylindrical with a 1.6 meter radius and a depth of 1 meter. The tanks are instrumented with a single PMT located at the top of the tank looking down into the water volume. On the inside, the tanks are lined with Tyvek$^{TM}$, a reflective material used to increase the light collection efficiency of the tanks.

The Milagro data used in this analysis were collected from July 2000 to March 2006 during which time the average duty factor was greater than 90%. The outrigger array was constructed and became operational in May 2003. Throughout this period the trigger rate of Milagro was ~1700 extensive air showers per second, most of which are due to cosmic ray showers incident on the ~2 sr field of view. The angular resolution of the Milagro data varies with number of PMTs hit and with the experimental configuration. On average, before the inclusion of the outrigger array the angular resolution was 0.75 degrees and after the inclusion of the outrigger array the average angular resolution was 0.5 degrees.

The background due to cosmic ray showers is estimated using the methods described in (Atkins *et al.* 2003): however, a more efficient parameter to distinguish between gamma rays and cosmic rays has been developed (Abdo 2006a & Abdo 2006b). This new parameter is especially useful for the data containing the array of outriggers. Known as $A_4$, the parameter is defined as

$$A_4 = \frac{(f_{Top} + f_{Out}) \times n_{Fit}}{PE_{Bot}^{max}},$$

where $f_{Top}$ and $f_{Out}$ are the fraction of PMTs in the top layer of the pond detector and the fraction of outrigger tanks hit respectively, $n_{Fit}$ is the number of PMTs retained in the angle fit and $PE_{Bot}^{max}$ is the number of photoelectrons in the bottom layer PMT with the most photoelectrons. Figure 1 shows the distributions of $A_4$ for proton and gamma ray Monte Carlo events and for data. The separation between gamma-ray events and background events is substantially improved especially for showers with cores that lie in the outrigger array and outside of the central reservoir. Such showers tend to be of higher energy than those with cores landing on the pond, so that the $A_4$ parameter improves the background rejection most at higher gamma-ray energies.

The signal to background (S/B) ratio is a strong function of $A_4$ as seen in Figure 1. As $A_4$ values increase, the probability that a shower is due to a gamma-ray, rather than a cosmic ray, increases. Therefore, instead of counting all the events in an angular bin equally, a weighted sum of events is used where events with higher values of $A_4$ are assigned higher weights. The significance is computed using the method Li and Ma (1983) where background fluctuations are similarly estimated for the sum of the weights of the background sample, rather than the background event count. The statistical significance derived from this analysis was verified both by Monte Carlo simulation and through the study of statistical fluctuations in the background data sample.

The values of the weights used in this analysis are determined from the predicted S/B ratio as calculated *a priori* from our detector simulation for an incident Crab-like spectrum. With this weighting scheme, this analysis is equivalent to a likelihood ratio method estimation in the limit that the background is large, which is true for the Milagro data. Different weights are used for the data with and without the outriggers and for different experimental configurations. The data is divided into 5 epochs for which different weights are calculated. In the data before the outriggers were installed, $f_{Out}$ is set equal to zero.

The source flux is determined by a forward modeling approach. In this method, the expected excess for an assumed source flux and spectrum is simulated for each interval in $A_4$. The expected counts in each interval are then multiplied by the weight determined from simulation for that interval. The results from each epoch are then multiplied by the duration of that epoch. The ratio of the observed excess to the sum of the weights from all epochs is then multiplied by the assumed source flux to get the observed flux. This method is repeated for all Declinations because the effective area of Milagro varies with zenith angle. In this manner the sensitivity of the results to different assumed source spectra can be investigated. The dependence of the derived source flux on the spectrum is minimized when it is quoted at the median detected energy, which is ~12 TeV for typical gamma-ray power law spectra and the weighted analysis using $A_4$. A change in the assumed source differential photon spectral index from -2.4 to -2.8 changes the quoted flux at 12 TeV by < 10%. With this technique, the Crab flux at 12 TeV is determined to be consistent with the flux measured by the HEGRA atmospheric Cherenkov telescope (Aharonian *et al.* 2004). However, the Milagro trigger rate as determined from simulations of protons and helium using the flux from direct measurements (Haino *et al.* 2004 & Asakimori *et al.* 1998) is underpredicted, so a systematic error of 30% is given to the gamma-ray fluxes quoted here.

The weighted analysis enhances the contribution of higher energy gamma rays. The median energy with this analysis is 12 TeV for a Crab-like spectrum of differential photon spectral index of -2.6. By comparison, previously published Milagro analysis had a median energy of 3-4 TeV. The combination of the installation of the outrigger array, the adoption of the $A_4$ discriminant, and the event weighting increases the sensitivity of Milagro as compared to previous analysis by ~2.5 times for a Crab-like spectrum and a greater factor for harder spectra.

**3. Results**

Figure 2 shows a TeV gamma-ray map of the northern hemisphere as determined from the Milagro data with this weighted analysis technique. The brightest extended region in the entire northern sky is the Cygnus Region of the Galactic plane located at roughly 21 hours in Right Ascension and 35 degrees in Declination. Figure 3 shows the TeV gamma-ray map of the Cygnus Region in Galactic coordinates. The contours in the figure show the matter density in the region. The TeV emission is correlated with the matter density with the exception of a significant deviation near Galactic longitude 75 degrees. This is the second brightest region of TeV emission (after the Crab Nebula) in the northern hemisphere and is observed with 10.9 standard deviations excess. This new source, MGRO J2019+37, is observed with a statistical significance ~ 6 standard deviations above the average diffuse gamma-ray emission in the region. The crosses in the figure show the location and location error of the EGRET sources (Hartman *et al.* 1999) in this region (none of which have been definitively identified with counterparts at other wavelengths).

The location of MGRO J2019+37 is R.A.=304.83 ± 0.14$_{stat}$ ± 0.3$_{sys}$ degrees and Dec.= 36.83 ± 0.08$_{stat}$ ± 0.25$_{sys}$ degrees. The systematic error is a combination of the 0.07 degree uncertainty in the Milagro location of the Crab, which is used to adjust the absolute pointing of Milagro and the uncertainties due to the unknown source morphology and to the diffuse background in the Cygnus region. While a definitive understanding of this new TeV source requires further multi-wavelength observations, the location of MGRO J2019+37 is consistent (within the combined location errors of EGRET and Milagro) with 2 EGRET sources. One of the EGRET sources (3EG

J2016+3657) is positionally coincident with the blazer-like source of unknown redshift, B2013+370 (Mukherjee *et al.* 2000) and the other (3EG J2021+3716) with the young pulsar wind nebula (PWN) G75.2+0.1 (Hessels *et al.* 2004, Roberts *et al.* 2002). An analysis of the highest energy photons (>1 GeV) observed by EGRET from this region indicates that the two sources were not resolved by EGRET. Given that the median energy of the Milagro observation is 12 TeV, a blazer-like source is less likely because such high energy gamma-rays are attenuated by interactions with the extragalactic infrared background.

While the angular resolution of Milagro for an average gamma ray is 0.5 degrees, the highest energy gamma rays detected have substantially better angular resolution (0.35 degrees) and background rejection. An examination of the arrival directions of the higher-energy photons shows that MGRO J2019+37 is most likely an extended source or multiple unresolved sources of TeV gamma rays. Fitting the source to a circular 2-D Gaussian, the width is σ=0.32±0.12 degrees. A comparison of the angular extent of the Crab and MGRO J2019+37 is shown in Figure 4. A fit with an elliptical 2-D Gaussian gives ~2 times larger extent in the direction of R.A. than Dec.

The distance to PWN G75.2+0.1 is estimated to be between 8-12 kpc from the dispersion measure of 369 pc cm$^{-3}$(Roberts *et al.* 2002). If MGRO J2019+37 is due to this PWN, then the source radius is 30-90 pc. However if the source lies within the Cygnus Region at a distance of 1-2 kpc, the source radius is only 4-15 pc. Assuming a differential source spectrum of E$^{-2.6}$, the Milagro flux measurement from a 3x3 square degree bin centered on the location given above for MGRO J2019+37 is given by E$^2$ dN/dE = (3.49 ± 0.47$_{stat}$ ± 1.05$_{sys}$) x 10$^{-12}$ TeV cm$^{-2}$ s$^{-1}$ at the median detected energy of 12 TeV. The diffuse flux from this region is difficult to determine, but as seen in Figure 3 it could be 30-40% of the total flux. EGRET measured the integral flux above 100 MeV of 3EG 2021+3716 to be (59.1 ± 6.2) x 10$^{-8}$ cm$^{-2}$ s$^{-1}$ with a differential spectral index at 100 MeV of -1.86 ± 0.10. The flux measured by Milagro above 12 TeV is a factor of 20-300 below an extrapolation of the EGRET spectrum (where the spread accounts for the errors on both the EGRET and Milagro measurements); therefore, if the two sources are the same, the spectrum must exhibit a spectral softening between 100 MeV and 12 TeV. A simple power-law fit between the 100 MeV flux and the 12 TeV flux (which includes the contribution of the diffuse flux in this 3x3 square degree bin) yields a differential photon spectral index of -2.22 ± 0.02.

The next brightest TeV region is just to the left of MGRO J2019+37 in Figure 3, at Galactic latitude of ~80$^o$, and is also coincident with an EGRET source (3EG J2033+4118) and the HEGRA source TeV J2032+413. The HEGRA source was detected between 1 and 10 TeV with a differential photon spectral index of -1.9 ± 0.1$_{stat}$ ± 0.3$_{sys}$, which when extrapolated to 12 TeV gives E$^2$ dN/dE = (7.9 ± 2.7 $_{stat}$) x 10$^{-13}$ TeV cm$^{-2}$ sec$^{-1}$ (Aharonian *et al.* 2005). The Milagro flux in a 3x3 square degree region centered on the HEGRA source at 12 TeV is (2.41 ± 0.48$_{stat}$ ± 072$_{sys}$) x 10$^{-12}$ cm$^{-2}$ s$^{-1}$ assuming a differential photon source spectrum of E$^{-2.6}$. Thus, the Milagro flux exceeds the HEGRA flux as is expected due to the additional contribution of the diffuse flux in this region. In fact, this region contains the largest matter density as can be seen from the contour lines of Figure 3. Further analysis is required to distinguish what fraction of the gamma rays observed by Milagro is from the HEGRA source or from the diffuse interactions.

To study the diffuse emission from the Cygnus region, a 3x3 degree square around MGRO J2019+37 is excluded from the area defined by Galactic latitude -3.0 to 3.0 degrees and Galactic longitude 65-85 degrees. For the remaining region, the energy flux at 12 TeV is $E^2 \, dN/dE = (4.18 \pm 0.52_{stat} \pm 1.26_{sys}) \times 10^{-10}$ TeV cm$^{-2}$ s$^{-1}$ sr$^{-1}$, assuming a differential photon source spectrum of $E^{-2.6}$.

## 4. Conclusions

The expected gamma-ray emissivity due to cosmic ray interactions with matter is predicted by the GALPROP (Strong *et al*. 2004a) program. The GALPROP model calculates the gamma-ray emissivities in every spatial grid point using the propagated spectra of cosmic ray species, leptons and nucleons, the interstellar radiation field, and the gas densities. The gas-related components (pion-decay and bremsstrahlung) of the gamma-ray sky maps are calculated using 21 cm line survey data for H I and CO J = 1 to J=0 survey data for $H_2$, in the form of column densities for Galactocentric rings, using velocity information and a rotation curve. The cosmic-ray source distribution is based on SNR/pulsars and a variable CO-to-$H_2$ conversion factor (Strong et al 2004b). In this way, details of Galactic structure are included.

The "conventional" model is tuned to have the propagated cosmic ray particle spectra and intensities match the local direct measurements. This model yields a deficit of diffuse gamma-ray emission above 1 GeV, a so-called GeV excess, observed in all directions on the sky. The "optimized" model (Strong *et al.* 2004a, 2004b) is tuned to match the EGRET diffuse emission data for the whole sky and reproduces the GeV excess by relaxing the constraints on matching the local cosmic ray proton and electron measurements. This "optimized" model is instead based upon the secondary antiprotons in cosmic rays and EGRET diffuse gamma-ray data. In this model the cosmic-ray intensities are significantly higher than those measured locally and the electron spectrum has been assumed to extend to well beyond 10 TeV to produce gamma rays in the Milagro energy range. The "optimized" model thus has a much larger contribution from inverse Compton scattering. Both models are shown in Figure 5 with the EGRET and Milagro measurements.

The Cygnus region is in a direction tangential to our spiral arm located at approximately the same distance from the Galactic center as the solar system. This direction is the most accurate to determine the gas distribution based on velocity information and the Galactic rotation curve. Therefore, the uncertainty in the determination of the gas distribution in this direction is minimal. As shown in Figure 5, the Milagro measurement of the diffuse flux in the Cygnus region is a factor of ~7 above predictions of the conventional model. This flux also exceeds the prediction of the optimized model which incorporates higher cosmic ray intensities to fit the EGRET data. Increasing the gas column density to agree with the Milagro data violate the restrictions imposed by the EGRET data. Both the parameters of GALPROP model and the Milagro flux measurement have large systematic uncertainties; however, the discrepancy between the model and the data likely imply the existence of an additional gamma-ray component. The spectrum of this component must be hard—for example, a differential photon spectral index of -2.3 to -2.4—in order to agree with fluxes measured by both the EGRET and Milagro.

There are several possible explanations for this component: unresolved sources of TeV gamma rays, a population of high-energy electrons in the region producing an inverse Compton flux at TeV energies, or a population of cosmic-ray accelerators which is dark because the hadrons do not interact near their sources but instead with the local matter distribution. The correlation of the observed emission with the matter density can be consistent with all of these explanations if the sources are co-located with the matter. If the excess is due to inverse Compton scattered photons, then this component must be a factor of ~20 higher at 12 TeV than the prediction of the conventional model.  Assuming a diffusion coefficient $D \sim 5.2 \times 10^{28}$ $(E/3 \text{ GeV})^{0.34}$ cm$^2$ s$^{-1}$ (Ptuskin *et al.* 2006) and characteristic energy loss timescale of $t \sim$ 30,000 yr (Strong *et al.* 1998), a 10 TeV electron travels a distance $x \sim (2Dt)^{0.5} \sim$ 200 pc.  Therefore, if the excess diffuse flux is due to inverse Compton scattered photons, the sources of the high-energy electrons must reside within the Cygnus region. The contribution of bremsstrahlung by cosmic ray electrons, another emission process that may correlate with the matter density, is negligible in the TeV energy range.  Also, the proposed explanation of the GeV excess due to neutralino annihilation (de Boer et al. 2005) cannot explain the Milagro high energy flux, because such a massive neutralino would have a much smaller number density and hence lower flux.   Therefore, this Milagro observation suggests that the Cygnus region contains hard-spectrum, cosmic-ray proton or electron accelerators.

**Acknowledgements**
The authors would like to thank Seth Digel for help in obtaining the HI and H$_2$ matter density contours.  This work has been supported by the National Science Foundation (under grants PHY-0245143, -0245234, -0302000, -0400424, -0504201, and ATM-0002744) the US Department of Energy (Office of High-Energy Physics and Office of Nuclear Physics), Los Alamos National Laboratory, the University of California, and the Institute of Geophysics and Planetary Physics.

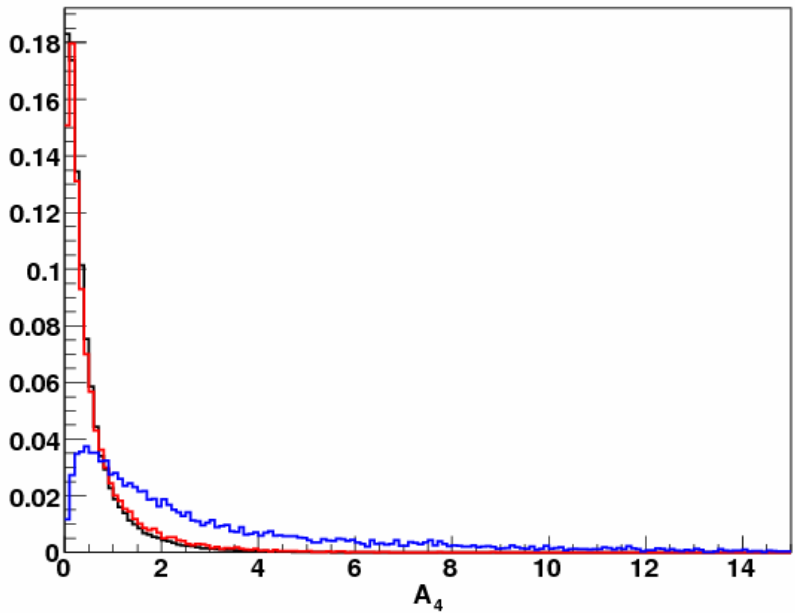

**Figure 1 | Distribution of $A_4$, a parameter used to distinguish the cosmic-ray background from the gamma-ray signal.** Monte Carlo predictions for proton initiated shower are shown in red and for gamma-ray initiated showers in blue. The y-axis is relative number of showers when the integral under each curve is equal to one. The differential particle spectrum for protons is chosen to be a power law of index -2.7 and for gamma-rays it is chosen to be -2.6. The $A_4$ distribution of the data (expected to be ~80% protons and the remainder heavier nuclei) is shown in black.

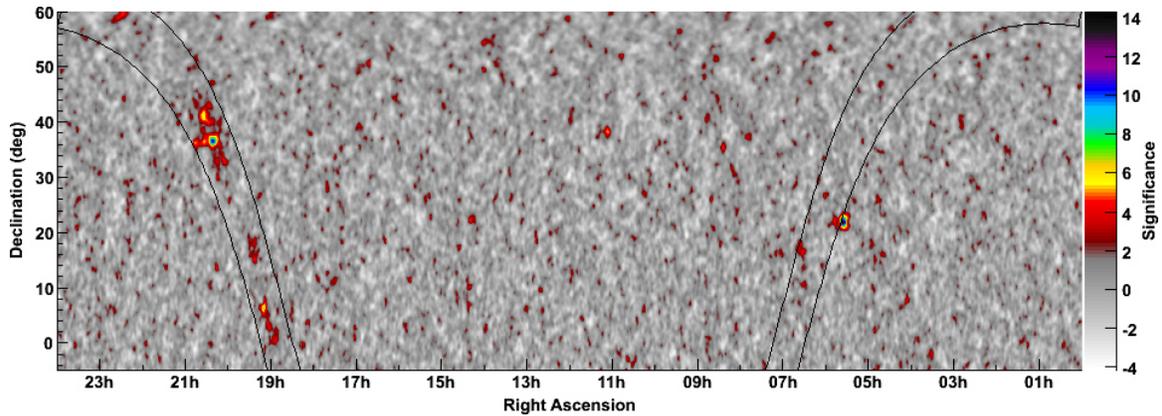

**Figure 2 | The northern hemisphere in TeV gamma rays.** The prominent feature towards the left of the figure is the Galactic Ridge. The Cygnus Region is within this ridge near a declination of 37 degrees. This map was made by smoothing each event by the point-spread function of the Milagro detector. At each point, the statistical significance of the observed excess (or deficit) is plotted. For clarity, locations with a statistical significance below 2 standard deviations are shown as monochrome. The dark lines show a ±5 degree region around the Galactic plane. The most significant point in the map is the Crab Nebula 14.2 σ located at a Right Ascension of 5h 34m and a Declination of 22 degrees.

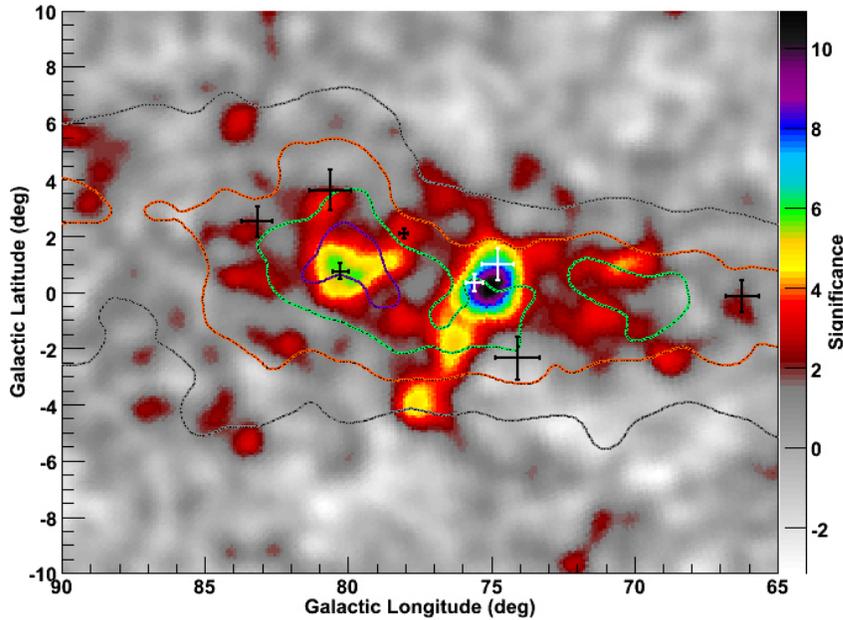

**Figure 3 The Cygnus Region of the Galaxy as seen in TeV gamma rays.** The color scale is the statistical significance of the gamma-ray excess at each location. Since the Milagro exposure and sensitivity are roughly constant over the region in the figure, the statistical significance is nearly proportional to the flux from each point. Superimposed on the image are contours showing the matter density in the region in steps of $3 \times 10^{21}/cm^2$ starting at $3 \times 10^{21}/cm^2$ with the outer grey contour. The matter density is a combination of atomic (HI) (Kalberla *et al.* 2005) and molecular hydrogen ($H_2$) (Dame *et al.* 2001). The 21 cm line survey data are used to determine the HI density and the CO J=1-0 survey data (Leung *et al.* 1992) are used to measure the $H_2$ density. The mass conversion factor $N_{H2}/W_{CO}$ is taken to be $X=0.8 \times 10^{20}$ $cm^{-2}$ $K^{-1}$ $km^{-1}$ s, which is the value used in this region in the GALPROP model (Strong *et al.* 2004b). With the exception of MGRO J2019+37 the matter density is correlated with the TeV excess, indicating that some of the TeV emission is due to interactions of cosmic rays with matter in the region. The crosses show the location of EGRET sources and their corresponding location errors.

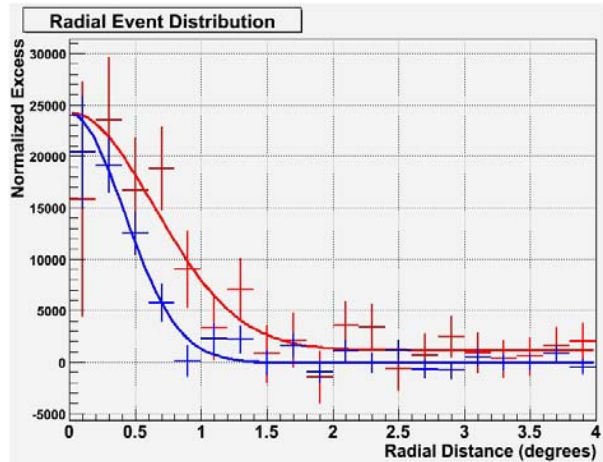

**Figure 4 Radial profile of events from the direction of the Crab (blue) and from MGRO J2019+37 (red).** Only events with large number of photomultiplier tube hits and outrigger information are shown due to their better angular resolution.

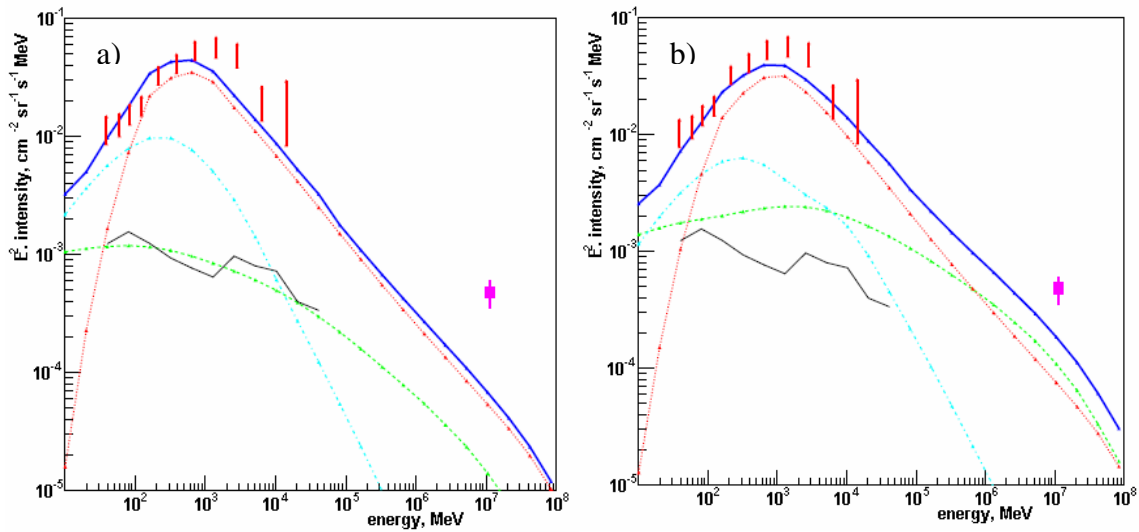

**Figure 5 Gamma-ray spectrum of the diffuse emission from the Cygnus Region of the Galactic Plane.** The red bars are the EGRET data and the purple bar is the Milagro measurement with the statistical error shown as a broad line and the systematic error shown as a narrow line. The other lines represent the different components of the emission according to the (a) "conventional" and (b) "optimized" GALPROP model of Strong *et al.* (2004a, 2004b). The solid blue line is the total predicted diffuse flux, the red line is the pion component, the green line is the component arising from inverse Compton interactions of high-energy electrons with optical and far infrared photons, the light blue line is due to proton bremsstrahlung, and the black line due to the extragalactic background. In the EGRET energy range, the model prediction includes the extragalactic background. Since the current Milagro analysis is insensitive to the isotropic, extragalactic, diffuse emission, the model prediction does not include this contribution in the Milagro energy range.